\documentclass[%
 reprint,
 amsmath,amssymb,
 aps,
]{revtex4-2}

\usepackage{graphicx}
\usepackage{dcolumn}
\usepackage{bm}

\begin{document}

\preprint{APS/123-QED}

\title{The solution to the Black Hole information paradox}

\author{Ivan Arraut}
\email{ivan.arraut@usj.edu.mo}
\affiliation{
Institute of Science and Environment and FBL,\\ 
University of Saint Joseph,\\
Estrada Marginal da Ilha Verde, 14-17, Macao, China
}




\date{\today}

\begin{abstract}
The information paradox suggests that the black hole loses information when it emits radiation. In this way, the spectrum of radiation corresponds to a mixed (non-pure) quantum state even if the internal state generating the black-hole is expected to be pure in essence. In this paper we propose an argument solving this paradox by understanding the process of spontaneous symmetry breaking when the black-hole selects one among the many possible ground states, emitting then radiation as a consequence of it. Here the particle operator number is the order parameter. This mechanism explains the connection between the density matrix corresponding to the pure state describing the black-hole state and the density matrix describing the spectrum of radiation (mixed quantum state). From this perspective, we can recover the black-hole information from the superposition principle applied to the different possible order parameters (particle number operators).
\end{abstract}

\maketitle


\section{Introduction}

The black-holes are compact objects containing a huge amount of mass in a very small spacetime region \cite{1, 111}. They develop a surface called event horizon. Once a particle enters the event horizon, classically it can never escape, no matter the energy invested on the process. Initially it was believed that the black-holes were completely black, unable to emit particles. However, it was subsequently demonstrated by Hawking that the black holes also emit radiation \cite{3, 4}. This result was important because without this emission of radiation, some fundamental laws of thermodynamics would be violated \cite{5, 6, 7}. The Hawking radiation however, brought by itself one problem, namely, the famous information paradox \cite{Inf}. The black hole information paradox was formulated by Hawking by the time he discovered the black-hole evaporation process. Hawking realized that while the radiation emerging from a black-hole is thermal in nature and it only depends on the mass, charge and angular momentum of the black hole, still there are infinite number of ways to generate same black hole with the same macroscopic properties. Then the same thermal radiation coming from the black-hole, could be developed by any of the infinite possible microstates (internal states) consistent with the black hole macrostate with mass $M$, charge $Q$ and angular momentum $J$. While each internal configuration of the black-hole represents a pure state, the thermal radiation, being associated with all the possible internal configurations, is represented by a mixed quantum state. This is basically the root of the information paradox. In this letter we propose an argument for solving the paradox. The starting point is the fact that the black hole evaporation can be expressed as a natural consequence of the spontaneous breaking of the symmetry under exchange of internal configurations. This symmetry keeps the same mass $M$, angular momentum $L$ and charge $Q$ invariant. We then demonstrate that the particle number corresponds to an order parameter. When the black hole selects one among the many possible ground states, it emits radiation and this is equivalent to tracing out all the other possible vacuum states. The difference between this case and the ordinary breaking of spontaneous symmetry breaking, is that here we have the possibility of having an entanglement between the different possible ground states. The information paradox disappears when we sum all the possible order parameters (particle number operators), showing that this sum must be equal to zero, recovering in this way the trivial ground state before the gravitational collapse. Then the original ground state, before the formation of the black-hole, is equal to the sum of all the possible ground states after the formation of the black-hole. We can interpret this as the possibility of emitting antiparticles in addition to the ordinary particles.

\section{The Schwarzschild solution}

For simplicity we will focus on the Schwarzschild case. For the other cases, the extension is direct. The metric is defined as \cite{Carrol}

\begin{equation}
ds^2=-\left(1-\frac{2GM}{r}\right)dt^2+\left(1-\frac{2GM}{r}\right)^{-1}dr^2+r^2d\Omega^2.    
\end{equation}
The event horizon is defined at $r_H=2GM$. Classically, when a particle approaches to a distance smaller than $r_H$, then it cannot escape from the gravitational influence of the black-hole. However, if we use arguments of Quantum Mechanics, then some particles can escape, giving rise to a spectrum of thermal radiation. The arguments developed by Hawking are explained in this section.

\subsection{The black hole evaporation process}

The black-hole evaporation process, emerges from a comparison between the vacuum state before the formation of the black hole and the vacuum state after the formation of the same body \cite{Cardip}. Before the formation of the black-hole, the vacuum state is trivial or devoid of particles, namely

\begin{equation}   \label{fvs}
\hat{b}_{\bf p}\vert\bar{0}>=0.    
\end{equation}
This vacuum state, corresponds to the field expansion

\begin{equation}   \label{field1}
\phi(x, t)=\sum_{\bf p}\left(f_{\bf p}\hat{b}_{\bf p}+\bar{f}_{\bf p}\hat{b}_{\bf p}^+\right).    
\end{equation}
After the formation of the black hole, the vacuum becomes nontrivial, and it is now defined by 

\begin{equation}   \label{fvs2}
\hat{a}_{\bf p}\vert0>=0.    
\end{equation}
This vacuum state then corresponds to the field expansion

\begin{equation}   \label{field2}
\phi(x, t)=\sum_{\bf p}\left(p_{\bf p}\hat{a}_{\bf p}+\bar{p}_{\bf p}\hat{a}_{\bf p}^++q_{\bf p}\hat{c}_{\bf p}+\bar{q}_{\bf p}\hat{c}_{\bf p}^+\right).
\end{equation}
It is important to remark that the field expansions in eq. (\ref{field1}) and (\ref{field2}), contain the same amount of information. The difference is with respect to which vacuum state we expand and as a consequence with respect to which modes we are expanding the quantum field. The effects of the radiation emerge when we compare the vacuum states of eqns. (\ref{fvs}) and (\ref{fvs2}). This comparison is possible via Bogoliubov transformations, able to relate the quantum operators as

\begin{equation}   \label{realbogo}
\hat{a}_{\bf p}=u_{{\bf p}, {\bf p'}}\hat{b}_{\bf p'}-v_{{\bf p}, {\bf p'}}\hat{b}_{\bf p'}^+. 
\end{equation}
Then when we try to annihilate the ground state defined in eq. (\ref{fvs}) with the operator $\hat{a}_{\bf p}$, the eq. (\ref{realbogo}) suggest 

\begin{equation}   \label{mama2}
<\bar{0}\vert\hat{n}_{\bf p}^a\vert\bar{0}>=\vert v_{{\bf p}, {\bf p'}}\vert^2.
\end{equation}
This means that the ground state now is full of particles, appearing through a spectrum of radiation. It has been proved before that 

\begin{equation}   \label{Statistics}
<\bar{0}\vert\hat{n}_{\bf p}^a\vert\bar{0}>=\frac{\Gamma_{{\bf p}, {\bf p'}}}{e^{\frac{2\pi\omega}{\kappa}}\pm1}.    
\end{equation}
The fact that the spectrum of radiation emerges from a non-zero value of the Bogoliubov coefficient $v_{{\bf p}, {\bf p'}}$, means that the Hawking radiation emerges from the mix of positive and negative frequency modes.  

\section{The formulation of the information paradox}

The no-hair theorem of black-holes suggests that the physical state of a black-hole, can be characterized by its mass $M$, angular momentum $L$ and charge $Q$ \cite{Carrol, Nohair}. These three independent parameters, are consistent with the huge amount of possible internal states of a black hole $\Omega$. Yet still, if the Hawking's calculation is right, then this means that the thermal spectrum is independent of the details about how the particles inside the black-hole are arranged \cite{Cardip}. Then technically the information about the internal details of the black hole is lost. Another way to perceive this, is by understanding that while each internal configuration of the black hole corresponds to a pure quantum state, the spectrum of radiation corresponds to a mixed quantum state. A pure quantum state, obeys a unitary evolution 

\begin{equation}
\vert\psi(t_1)>=U(t_1, t_2)\vert\psi(t_2)>,
\end{equation}
and we can always express it as a ket (wave function). More generally, it is also possible to express the quantum state through a density matrix as  $\hat{\rho}=\vert\psi><\psi\vert$. For pure and mixed states the trace of this operator is $Tr(\hat{\rho})=1$. However, although for pure states the idempotent condition $\hat{\rho}^2=\hat{\rho}$, or equivalently, $Tr(\hat{\rho}^2)=1$; for mixed states this condition is violated and in general $Tr(\hat{\rho}^2)\leq1$. Although there is no ket representation for mixed states, still they can be expressed with a density matrix of the form

\begin{equation}   \label{Mixing}
\hat{\rho}=\sum_{k=1}^Np_k\vert\psi_k><\psi_k\vert.
\end{equation}
Here $\vert\psi_k>$ is some set of pure states. The eq. (\ref{Mixing}) is then a superposition of pure states. The essential idea behind the information paradox is that the black-hole radiation is thermal because it corresponds to a mixed quantum state, which is a superposition of the pure states represented by all the possible internal configuration arrangements of the black hole. 

\section{The information paradox from the perspective of spontaneous symmetry breaking}

\begin{figure}
	\centering
		\includegraphics[width=0.4\textwidth]{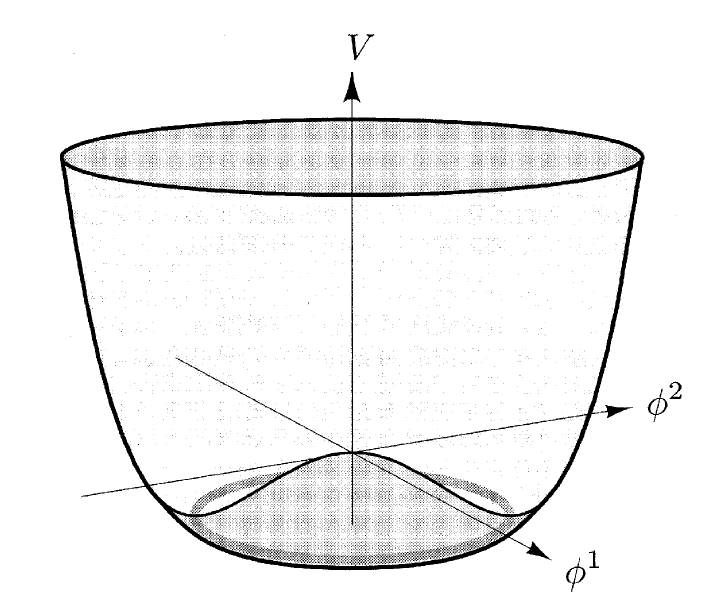}
	\caption{The typical "Mexican hat" potential generated when we have degenerate vacuum states. Each point at the bottom of the potential represents a vacuum state. When the black hole selects one state, it breaks the symmetry spontaneously and then it emits radiation. Figure taken from \cite{Peskin}.}
	\label{Fig.1}
\end{figure}

The black-hole evaporation process, can be expressed as a consequence of the mechanism of spontaneous symmetry breaking where the black-hole, having access to a huge amount of possible ground states, selects one arbitrarily. When this occurs, then the black-hole emits particles in the form of radiation. Each different internal configuration corresponds to a different ground state. In principle, if we follow the standard approaches, then the different vacuum states would correspond to different Hilbert spaces at the thermodynamic limit \cite{Br}. Yet still, we will see later that the ground states are entangled to each other. The Lagrangian governing the dynamics of the emitted particles is

\begin{equation}   \label{KGextended}
\pounds=\frac{1}{2}\partial^\mu\hat{n}^a_{\bf p}(\omega)\partial_\mu\hat{n}^a_{\bf p}(\omega)-V(\hat{n}^a(\omega)).
\end{equation}
The potential $V(\hat{n}_{\bf p})$, is defined as 

\begin{equation}   \label{KGextended2}
V(\hat{n}_{\bf p})=\frac{1}{2}m^2\hat{n}_{\bf p}^2+\frac{\beta}{3}\hat{n}^3_{\bf p}+\frac{\lambda}{4}\hat{n}^4_{\bf p}.
\end{equation}
We could calculate the ground state by using the condition $\partial V/\partial\hat{n}=0$. However, due to the spacetime curvature, still some kinetic term remains and it cannot be ignored in eq. (\ref{KGextended}) and then we have to consider the full version of the Euler-Lagrange equations. The symmetry of the system under exchange of internal configurations, consistent with the Black-Hole entropy (exchange of particles), is spontaneously broken when $m^2<0$. The signature of the parameter $\beta$ tells us whether the particles evaporating are bosons or fermions. The spectrum of radiation emerges from applying the Euler-Lagrenge equations over the Lagrangian (\ref{KGextended}). The information paradox from this perspective suggests that while the trivial ground state, before the formation of the black-hole, can be represented with a pure quantum state; after the formation of the black-hole, the vacuum is degenerate with each vacuum state being represented by each point at the bottom of the potential on the figure (\ref{Fig.1}). Each possible ground state makes a thermal emission of particles if the black-hole selects them during the process. The thermal emission then corresponds to a mixed quantum state while the quantum state corresponding to the internal configuration of the Black-Hole is supposed to be a pure quantum state. This is the source of the information paradox. In the coming section we will see how to solve this important problem.

\section{Mixed quantum states and the degenerate vacuum}

The fact that the evaporation process of a black-hole corresponds to a spectrum of radiation, means that the emitted radiation is independent of the possible internal configurations of the black-hole. The degeneracy of the ground state is consistent with this statement. However, more generally, this means that all the possible internal configurations are entangled with each other. Initially we would be tempted to express the superposition of each possible internal configuration of a black hole with mass $M$, charge $Q$ and angular momentum $L$ as

\begin{equation}   \label{AmazingFrog1}
\vert\psi>=\frac{\vert0>_1+\vert\psi>_2+\vert0>_3+...+\vert0>_N}{\sqrt{N}}.
\end{equation}
Here $\vert0>_i$ corresponds to each possible ground state of the black-hole. The quantum state (\ref{AmazingFrog1}) corresponds to a pure state as it can be proved if we construct the density matrix. The only problem with this state, is that it suggests that all the possible internal ground states are not entangled. In order to represent the entanglement of the different internal configurations, it is appropriate to express the system as a quantum state of the form 

\begin{eqnarray}   \label{AmazingFrog2}
\vert\psi>=\frac{\vert1>_1\bigotimes\vert0>_2\bigotimes\vert0>_3\bigotimes...\bigotimes\vert0>_N}{\sqrt{N}}\nonumber\\
+\frac{\vert0>_1\bigotimes\vert1>_2\bigotimes\vert0>_3\bigotimes...\bigotimes\vert0>_N}{\sqrt{N}}\nonumber\\
+\frac{\vert0>_1\bigotimes\vert0>_2\bigotimes\vert1>_3\bigotimes...\bigotimes\vert0>_N}{\sqrt{N}}\nonumber\\
+\frac{\vert0>_1\bigotimes\vert0>_2\bigotimes\vert0>_3\bigotimes...\bigotimes\vert1>_N}{\sqrt{N}}.
\end{eqnarray}
This quantum state, looks more like the state that Hawking imagined, considering that the thermal radiation has to come from tracing out all the possible ground states, except the one which the black-hole selects when it breaks the symmetry of the system spontaneously. Expressing the black-hole ground state as in eq. (\ref{AmazingFrog1}) or (\ref{AmazingFrog2}) is a matter of convention and it will not affect the conclusions. We just have to keep in mind that the different ground states are entangled. The pure quantum state represented by the state (\ref{AmazingFrog2}) can be expressed through the density matrix

\begin{equation}   \label{Amazing3}
\hat{\rho}=\left(\begin{array}{cccc}
   \frac{1}{N} &\frac{1}{N} & ... & \frac{1}{N}\\
   \frac{1}{N} & \frac{1}{N}&...&\frac{1}{N}\\
   ...&... & ... & ... \\
   ... & ... &\frac{1}{N} & \frac{1}{N}
\end{array}\right).
\end{equation}
We can verify that $Tr(\hat{\rho})=1$ (the trace runs over $N$ entries of the matrix) and that the density matrix is idempotent, which means that the ground state of a black-hole, before breaking the symmetry under exchange of particles, is a pure state. Breaking spontaneously the symmetry under exchange of configurations, is equivalent to tracing out all the ground states in eq. (\ref{Amazing3}), except the one which the black hole selects. Evidently, any selected ground state, after tracing out all the additional ground states, is a mixed quantum state (non-pure). This is what gives a thermal character to the spectrum of radiation emitted by the black-hole. For solving the information paradox, we can argue that each possible ground state corresponds to a mixed quantum state with probability $p_k=1/N$ as it is formulated in agreement with eq. (\ref{Mixing}). Here $N$ accounts the number of possible ground states, while each ground state has equal probability to appear. The states are entangled with each other. The selection of a specific ground state, depends on external factors like the direction of expansion of the universe for example. In this way, we can describe the evaporation process of a black-hole as follows: 1). The black-hole is formed and characterized by its mass $M$, angular momentum $L$ and charge $Q$. 2). A degenerate ground state emerges as a consequence of all the internal states consistent with the specific values taken by $M$, $L$ and $Q$. 3). Each ground state has equal probability $p_k=1/N$ to emerge as the final ground state. 4). Each ground state is not a pure quantum state but the total wave function of the system is a superposition of all the possible ground states, giving then a final pure (combined) ground state. 5). The selection of a single ground state, naturally gives a thermal spectrum corresponding to a mixed quantum state. If we trace out all the ground states, except one, then the resulting density matrix is

\begin{equation}   \label{Amazing4}
\hat{\rho}_i=Tr(\hat{\rho})_{j\neq i}=\left(\begin{array}{cccc}
   \frac{1}{N} &0 & ... & 0\\
   0 & \frac{1}{N}&...&0\\
   ...&... & ... & ... \\
   ... & ... &0& \frac{1}{N}
\end{array}\right).
\end{equation}
This density matrix naturally represents a mixed quantum state. It is easy to verify that $Tr(\hat{\rho}_i)=1$, $\hat{\rho}_i^2\neq\rho_i$ and then $Tr(\hat{\rho}_i^2)<1$. At this point it is clear that while the black-hole quantum state is a pure state as the density matrix (\ref{Amazing3}), still the Quantum state representing the thermal spectrum, and defined in eq. (\ref{Amazing4}), 

\subsection{Recovering unitarity}

It is possible to recover the unitarity of the system if we make a superposition of all the possible spectrum of radiation emitted by the black hole. In other words, we can recover the original ground state if we take the particle number operator $\hat{n}_i$ as the order parameter as it was suggested in \cite{Mypaper}. Before the formation of the black-hole, $<\bar{0}\vert\hat{n}_{\bf p}\vert\bar{0}>=0$, while after the formation of the black-hole we have $<\bar{0}\vert\hat{n}_{\bf p}\vert\bar{0}>\neq0$. Considering the vacuum degeneracy, if we sum all the possible values of the order parameter, we should recover the trivial result \cite{M2, M3, M4, M5, M6}. In this way, we can say that 

\begin{equation}   \label{Solved}
<0\vert\hat{n}_1\vert0>_1+<0\vert\hat{n}_2\vert0>_2+...+<0\vert\hat{n}_N\vert0>_N=0.    
\end{equation}
This means that depending on the external perturbation breaking the symmetry of the black-hole, the particle number can take positive and negative values after evaluating the corresponding vacuum expectation values. This is only possible if the black-hole emits particles in some situations and antiparticles in others, taking then the emission of antiparticles as a negative number operator after evaluating the corresponding expectation value. In this way, in general the black-holes can emit particles and antiparticles at any instant. This solves the paradox, because eq. (\ref{Solved}) is telling us that 

\begin{equation}   \label{Solved2}
<\bar{0}\vert\hat{n}_1\vert\bar{0}>_1+...+<\bar{0}\vert\hat{n}_N\vert\bar{0}>_N=<0\vert\hat{n}_{bf}\vert0>=0,    
\end{equation}
where the subindex $bf$ means "before formation" of the black-hole. Then it is possible to recover all the information of the black-hole by summing all the possible outcomes of the order parameter evaluated at their corresponding ground states, taking the order parameter as the particle number operator. When the emitted field represents a particle, then $<\bar{0}\vert\hat{n}_i\vert\bar{0}>_i>0$ and when the emitted field represents an antiparticle $<\bar{0}\vert\hat{n}_i\vert\bar{0}>_i<0$. This means that half of the degenerate ground states ($N/2$) correspond to the emission of particles, while the other half ($N/2$) represent antiparticles.

\section{Conclusions}

In this paper we have demonstrated that it is possible to solve the black-hole information paradox if we consider the process as a consequence of breaking the symmetry under internal exchange of configurations spontaneously. From this perspective, the particle operator number $\hat{n}_{\bf p}$ is the order parameter of the system, which is trivial before the formation of the black hole (zero vacuum expectation value) and non-trivial after the formation of the black-hole (non-zero vacuum expectation value), once one of the degenerate vacuum states is selected during the process. Yet still, the whole quantum state representing the black-hole, corresponds to a pure quantum state which considers all the possible (degenerate) vacuum states. Selecting one of the ground states is equivalent to tracing out all the other vacuum states and this is why the Hawking radiation has a thermal nature, generating then a mixed quantum state (non-pure). The information of the black-hole is recovered when we sum all the possible order parameters emerging from breaking the symmetry, under internal configurations, spontaneously. This brings another physical consequence suggesting that the black-holes not only emit particles but also antiparticles. The present formulation solves the information paradox in black-holes.

\end{document}